\documentclass[acus]{JAC2003}

\usepackage{graphicx}
\usepackage{booktabs}

\begin{document}
\title{BEAM DYNAMICS STUDIES FOR THE HIE-ISOLDE LINAC AT CERN\thanks{The author (M.A. Fraser) acknowledges the receipt of funding from the ISOLDE Collaboration Committee and the Cockcroft Institute.}}
\author{M.A. Fraser$^{\dag \ddag}$, M. Pasini$^{\S \ddag}$, R.M. Jones$^{\dag}$, M. Lindroos$^{\ddag}$\\
$^{\dag}$The University of Manchester, Oxford Road, Manchester, M13 9PL, UK.\\
$^{\dag}$The Cockcroft Institute, Daresbury, Warrington, Cheshire WA4 4AD, UK.\\
$^{\S}$Instituut voor Kern- en Stralingsfysica, K.U.Leuven, Celestijnenlaan 200D
B-3001 Leuven, BE.\\
$^{\ddag}$CERN, Geneva, Switzerland.\\}

\maketitle

\begin{abstract}
   The upgrade of the normal conducting (NC) Radioactive Ion Beam EXperiment (REX)-ISOLDE heavy ion accelerator at CERN, under the High Intensity and Energy (HIE)-ISOLDE framework, proposes the use of superconducting (SC) quarter-wave resonators (QWRs) to increase the energy capability of the facility from 3 MeV/u to beyond 10 MeV/u. A beam dynamics study of a lattice design comprising SC QWRs and SC solenoids has confirmed the design's ability to accelerate ions, with a mass-to-charge ratio in the range 2.5 $<$ A/q $<$ 4.5, to the target energy with a minimal emittance increase. We report on the development of this study to include the implementation of realistic fields within the QWRs and solenoids. A preliminary error study is presented in order to constrain tolerances on the manufacturing and alignment of the linac.
\end{abstract}

\section{INTRODUCTION}

\begin{figure}[htb]
   \centering
   \includegraphics*[width=82mm]{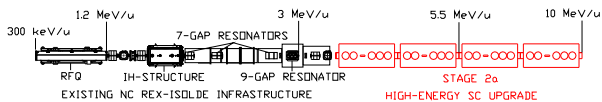}
   \caption{The stage 2a layout of the HIE-LINAC.}
   \label{stage2a}
\end{figure}

The HIE-LINAC strives to increase the energy and quality of post-accelerated radioactive ion beams (RIBs) delivered by the ISOLDE nuclear facility at CERN~\cite{tech_opts}. The linac will comprise of two dedicated sections for low and high energy, containing cavities designed with geometric reduced velocities, $\beta_{0}$, of 6.3\% and 10.3\% respectively. The high energy section will be built first and boost the energy of the existing facility from 3~MeV/u to 10~MeV/u. The low energy section will replace the NC 7-gap and 9-gap resonators and provide full energy variability of the RIB. We present studies of the high-$\beta$ cavity and full beam dynamics simulations of stage 2a of the HIE-LINAC, which is outlined schematically in Fig.~\ref{stage2a}.

\begin{figure}[htb]
   \centering
   \includegraphics*[width=65mm]{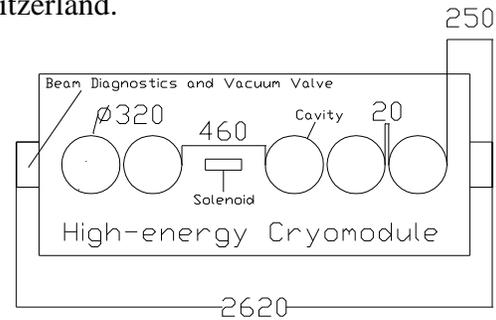}
   \caption{The lattice design of the high-energy cryomodule. Dimensions in~mm.}
   \label{high_cryo}
\end{figure}

The design of the HIE-LINAC lattice has been discussed in~\cite{bd_LINAC08} and the high energy cryomodule lattice is presented schematically in Fig.~\ref{high_cryo}.

The design of the solenoid is being developed at CERN~\cite{private_remo}, with the capability of delivering a longitudinal axial peak field of 120~kG and a stray field at the adjacent cavities of only 0.02~G. A three dimensional particle tracking routine has been written in order to study the single particle dynamics within the high-$\beta$ cavity.  The electromagnetic cavity fields implemented in the tracking routine were simulated using the \texttt{MWS}~\cite{MWS} code. The multi-particle beam dynamics simulations were performed by implementing the field maps in the program \texttt{TRACK}~\cite{TRACK}.

\section{SINGLE PARTICLE BEAM DYNAMICS IN THE HIGH-$\beta$ CAVITY}

The SC QWRs operate at a frequency of 101.28~MHz and are designed to deliver a gradient of 6~MV/m. They posses two gaps and operate in $\pi$-mode, providing efficient acceleration over a wide velocity range. The cavities are independently phased and able to deliver beams of optimal energy for the specified range of mass-to-charge states. More details about the cavity design can be found in~\cite{cav}. All stated results consider the QWRs accelerating the beam at a synchronous phase of -20 degrees with respect to the phase of maximum energy gain.

\begin{figure}[htb]
   \centering
   \includegraphics*[width=80mm]{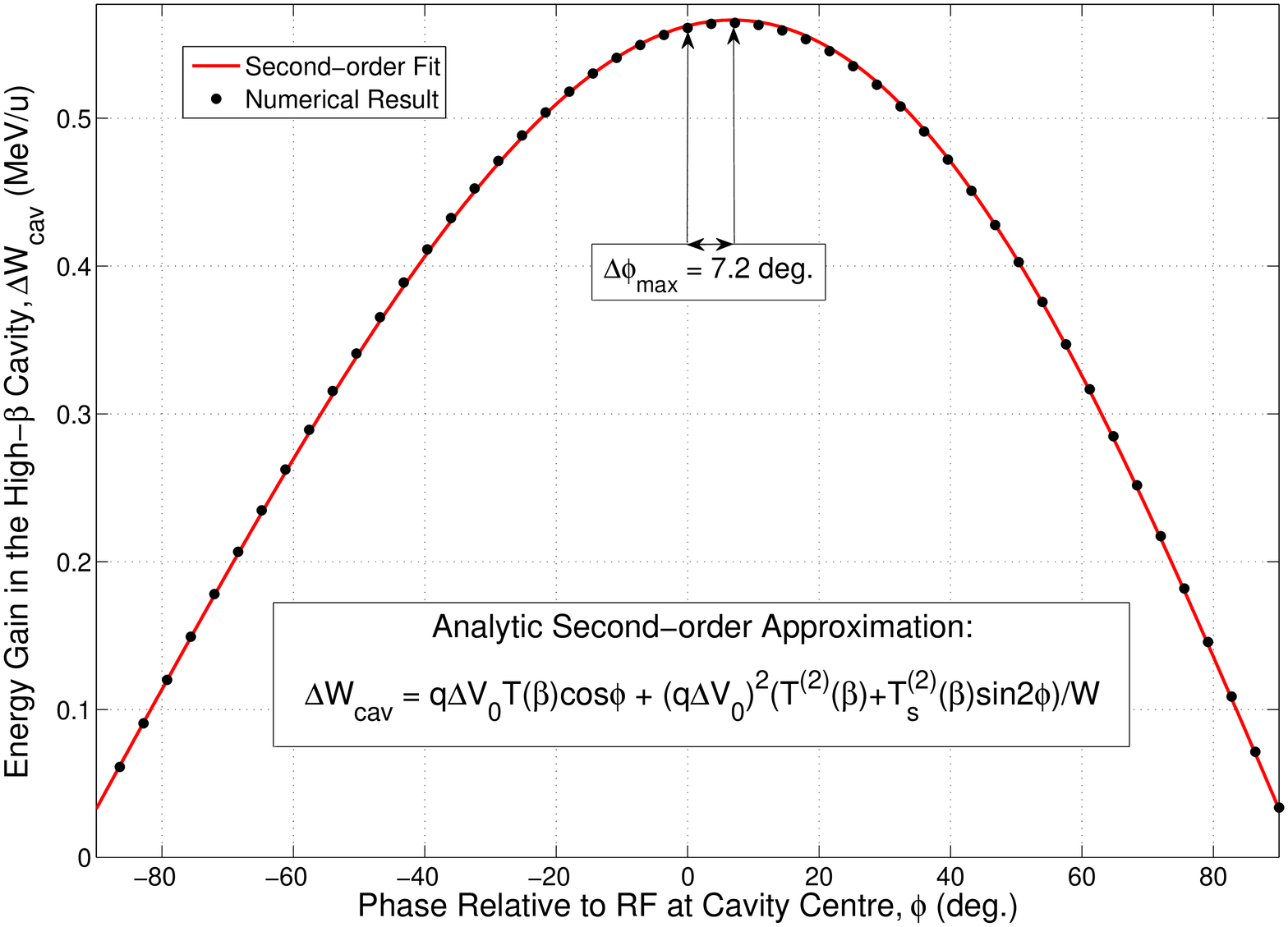}
   \caption{The energy gain as a function of phase for ions with A/q = 2.5, injected into the high-$\beta$ cavity at 8\% the speed of light.}
   \label{delayen}
\end{figure}

The energy gain on axis was calculated numerically and compared to an approximation in which the accelerating field is constant within the gaps and the ion transit velocity through the cavity is constant. This first-order approximation is good to better that 1\% for velocities above $\beta_{0}$, however, a discrepancy of more than 5\% is observed for the lightest ions at injection energy. In this case, the velocity change in the cavity can reach almost 10\% and in order to quickly and accurately predict the energy gain we are forced to use a second-order analytic approximation, as derived by Delayen in~\cite{delayen}. The analytic expression is shown in Fig.~\ref{delayen}, where all the symbols have the same meaning as in the quoted reference: $T$ is the first-order transit time factor (TTF) and $T^{(2)}$ and $T^{(2)}_{s}$ are the second-order TTFs. The expression allows for a quick computation of the energy gain to better than 0.5\% for ions entering the cavity with any phase and reduced velocities as low as 7.8\% of the speed of light. 

As a result of the change in velocity, lighter ions at injection energy don't receive maximum acceleration by crossing the cavity centre when the electromagnetic fields are zero and reversing. In the extreme case, with a beam of A/q = 2.5 at 3 MeV/u, we calculate that the maximum energy gain occurs for ions that cross the cavity centre 7.2 degrees in phase later than the switch of sign of the fields. In order to maintain the correct synchronous phase and design energy the bunches must be shifted accordingly in phase.

\section{COMPENSATION OF BEAM STEERING}

\begin{figure}[htb]
   \centering
   \includegraphics*[width=80mm]{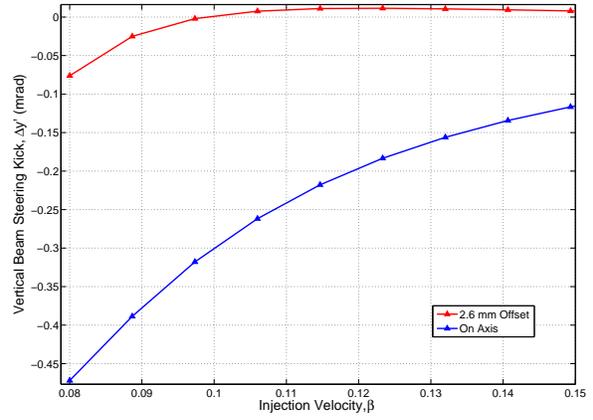}
   \caption{The compensation of a beam with A/q = 4.5.}
   \label{comp}
\end{figure}

The presence of a non-negligible horizontal magnetic field component on the axis of the QWR necessitates a compensation scheme in order to reduce the kick on the beam within the cavity. The beam steering effect is strong and has a magnitude of several tenths of milliradians at the cavity axis. The phase dependency of the steering force can couple the longitudinal and transverse motions causing emittance growth and it is for this reason that we consider a compensation scheme within the cavity, in addition to correctors placed outside of the cryomodules. We employ the compensation scheme described by Ostroumov in~\cite{beam_steer} and deliberately offset the beam axis vertically by 2.6 mm in the cavity. This introduces a compensating electric field component which creates a force in opposition to the magnetic steering force acting on the beam. The amount of offset was chosen by minimising the integrated beam steering effect along the linac and selecting the best compromise for the design specification of mass-to-charge states. After compensation, the beam steering effect is significantly reduced over a broad velocity range as shown in Fig.~\ref{comp}.

\begin{figure}[htb]
   \centering
   \includegraphics*[width=80mm]{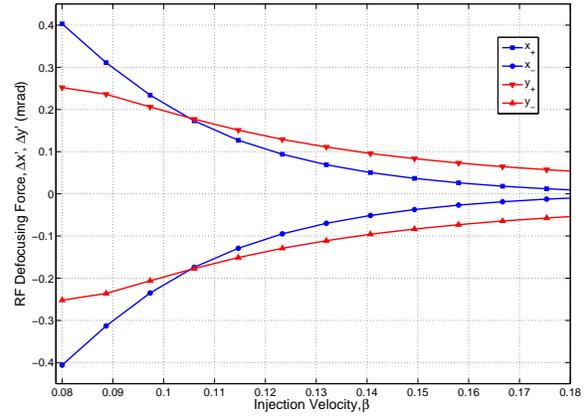}
   \caption{The defocusing kick on ions of A/q = 2.5 offset 1~mm from the axis in the horizontal and vertical directions.}
   \label{RF_defocus}
\end{figure}

The difference in the velocity dependence of the electric and magnetic forces causes a deterioration in the compensation quality at low velocity. The RF defocusing forces in the horizontal, \emph{x}, and vertical, \emph{y}, directions at 1 mm either side of the offset beam axis are shown in the Fig.~\ref{RF_defocus}. The asymmetry of the RF defocusing force has consequences for the transverse beam emittance when the beam is rotated in the solenoid focusing channel. The emittance growth of the beam in the solenoid focusing channel is studied in the realistic field beam dynamic simulations.

\section{REALISTIC FIELD BEAM DYNAMICS SIMULATIONS}

\begin{figure}[tb]
    \centering
    \includegraphics[width=80mm]{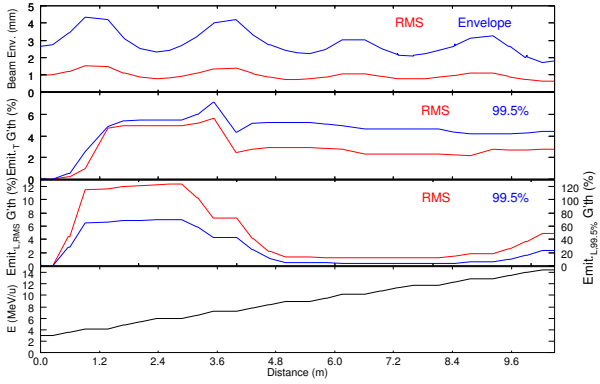}
    \caption{The beam dynamics with a RIB of A/q = 2.5.}
    \label{stage2a_bd_2_5}
\end{figure}

\begin{figure}[tb]
    \centering
    \includegraphics[width=80mm]{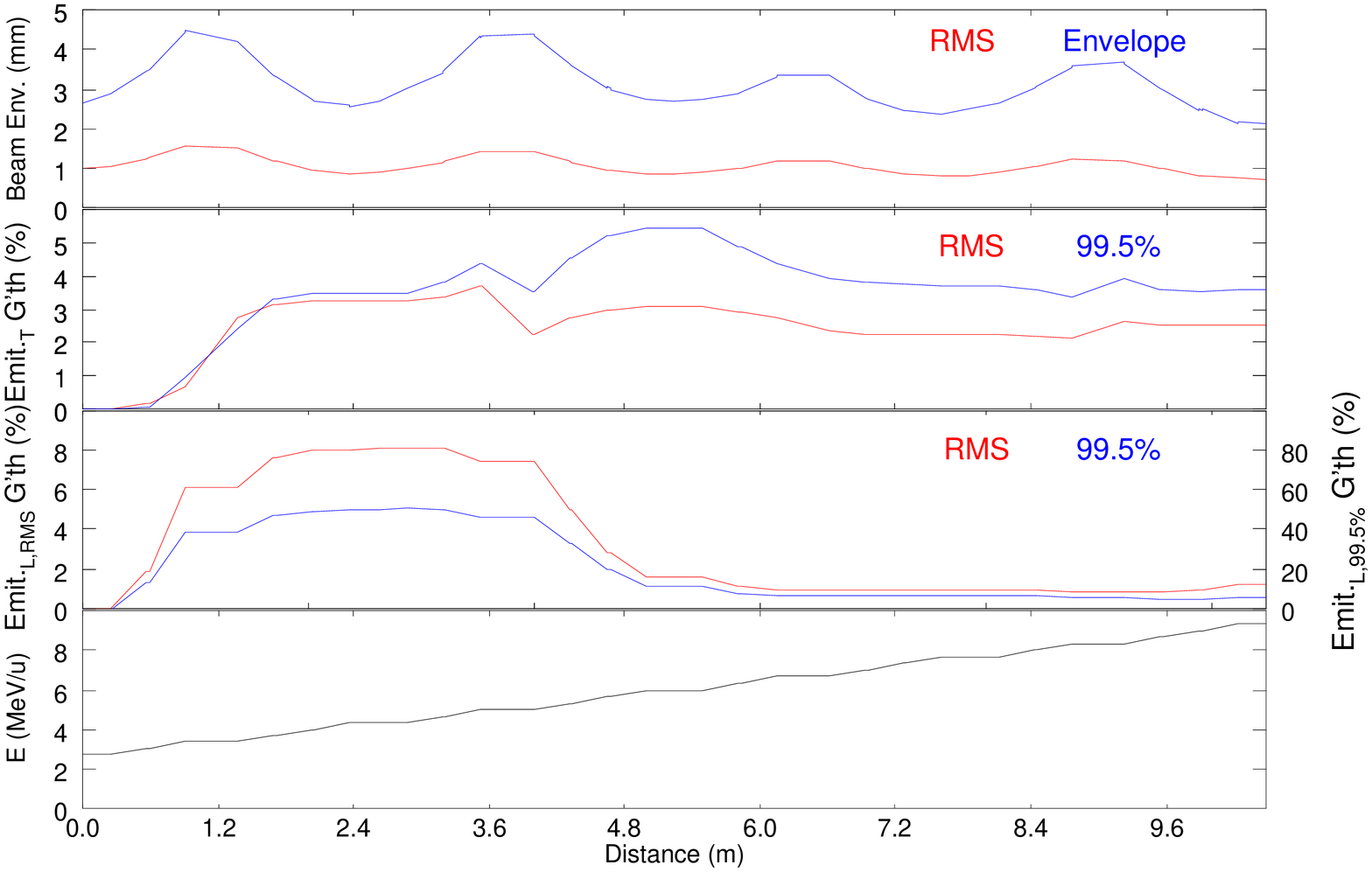}
    \caption{The beam dynamics with a RIB of A/q = 4.5.}
    \label{stage2a_bd_4_5}
\end{figure}

The latest electromagnetic field maps were implemented into the \texttt{TRACK} code in order to run multi-particle simulations and study the evolution of the beam emittance along stage 2a of the HIE-LINAC.  Space charge forces are neglected because of the low beam intensity. In Figs.~\ref{stage2a_bd_2_5} and~\ref{stage2a_bd_4_5}, we present the simulation results of 50000 macro-particles generated with a 6D waterbag distribution and with A/q = 2.5 and 4.5, entering an aligned linac operating at 90 degrees transverse phase advance per period. The transverse emittance growth of the beam is minimal and its source is dominated by the phase spread of the beam at injection in the QWRs of the first cryomodule. The emittance growth arising from the asymmetric transverse distortion of the beam coupled with rotation in the solenoids contributes less than 1\% to the RMS emittance growth. The beam parameters are summarised in Table~\ref{bp}.

\begin{table}[hbt]
   \centering
   \small
   \caption{Beam Parameters for Stage 2a of the HIE-LINAC}
   \begin{tabular}{lcc}
       \toprule
       \textbf{Parameter (A/q)} & \textbf{Input (2.5/4.5)} & \textbf{Output (2.5/4.5)} \\ 
       \midrule
           $\alpha_{t}$     & -0.05/-0.15             & -0.1/-0.21             \\
           $\beta_{t}$ (mm/mrad)        & 1.0/1.0            & 1.0/1.0          \\
            $\epsilon^{99.5\%}_{t,norm}$ ($\pi$ mm mrad)        & 0.25/0.25             & 0.28/0.26       \\
            $\alpha_{l}$         & 1.37/1.37  & -1.05/0.34        \\
           $\beta_{l}$ (deg/keV)        & 0.043/0.026            &0.015/0.012       \\
            $\epsilon^{99.5\%}_{l,norm}$ ($\pi$ ns keV/u)        & 1.66/1.66           & 2.05/1.74       \\
            Energy (MeV/u) & 3.07/2.80 & 14.49/9.31 \\
         \bottomrule
   \end{tabular}
   \label{bp}
\end{table}

\section{MISALIGNMENT STUDY}

\begin{figure}[tb]
    \centering
    \includegraphics[width=80mm]{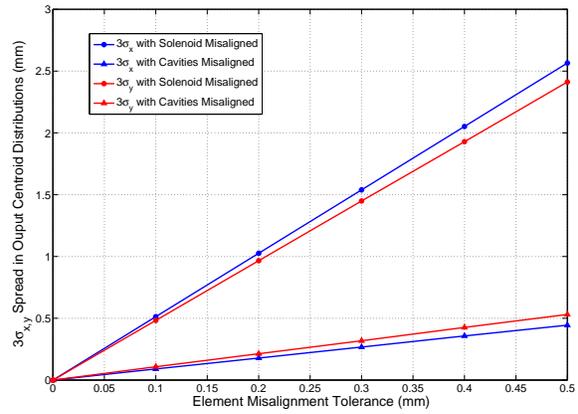}
    \caption{The transverse spread of the beam centroid distribution resulting from random misalignments of either the solenoid or the cavities in the HIE-LINAC.}
    \label{error_study}
\end{figure}

A misalignment study is being carried out in order to specify the tolerances on the alignment of the elements based on the demands of the beam dynamics. The elements will be aligned to an internal reference within each cryomodule and each cryomodule will be aligned separately with the beam. For this reason we first consider a misalignment study of the elements within an individual cryomodule. We use the \texttt{TRACK} code to randomly misalign elements within the first cryomodule in the transverse plane. The procedure simulates the beam centroid as a paraxial single particle at injection and no correction routine is applied. The code is looped over 1000 randomly misaligned linacs within a given tolerance and the output distribution of centroids in phase-space is statistically analysed. The misalignment tolerances are symmetric in the horizontal and vertical transverse directions. The distribution of centroids at output from the first cryomodule is Gaussian and closely independent of the mass-to-charge state of the ion. We present the results of separately misaligning the solenoid and cavities in Fig.~\ref{error_study}. The tolerance on the alignment of the solenoid in the transverse plane must be on average 5.7 times more stringent than the cavities in order to attain the same RMS spread in the centroid distribution after the first cryomodule. The solenoid will require a separate and specialised alignment system within the cryomodule in order to cope with a greater demand for alignment.

\end{document}